\documentclass[fleqn,pdflatex]{article-hermes}

\journal{\LOBJET. Volume 8 -- n2/2005}{1}{15}

\publisher{00 00 00 00 00 }{00 00 00 00 00}{guillaume.laurent@ens2m.fr}

\title[Mode d'emploi de \textit{article-hermes.cls}]%
      {Common Proper Motion Search for\\ Faint Companions
       Around Early-Type\\ Field Stars - Progress Report - II.}

\subtitle{}
\author{Valentin D. Ivanov\fup{*}
   \andauthor D. N\"urnberger\fup{*}
   \andauthor G. Chauvin\fup{*}
   \andauthor\\ C. Foellmi\fup{*}
   \andauthor M. Hartung\fup{*}
   \andauthor N. Hu\'elamo\fup{**}
   \andauthor C. Melo\fup{*}
   \andauthor\\ M. Sterzik\fup{*}
   \andauthor X. Haubois\fup{***}}

\address{%
\fup{*} European Southern Observatory,
Ave. Alonso de Cordova 3107
Vitacura, Casilla 19001, Santiago 19, Chile\\
[3pt] vivanov,dnuernbe,gchauvin,cfoellmi,mhartung,cmelo,msterzik@eso.org\\
\fup{**} Observat\'orio Astron\,omico de Lisboa,
Tapada da Ajuda - 1349-018 Lisboa, Portugal\\
[3pt] nhuelamo@oal.ul.pt\\
\fup{***} Observatoire de Paris, Place Jules Janssen, F-92 195
Meudon Cedex, France\\
[3pt] xavier.haubois@obspm.fr
}

\resume{
La multiplicit\'e des \'etoiles pr\'ecoces n'est aujourd'hui
toujours pas bien \'etablie. Les \'etudes de r\'egions de
formation stellaire individuelles sugg\'erent une relation avec
l'\^age et l'environnement. Afin de r\'epondre \`a ces
questions, nous avons engag\'e le premier relev\'e d\'etaill\'e
d'imagerie assist\'ee par optique adaptative sur 308 \'etoiles
du champ de type BA, dans un rayon de 300\,parsecs autour du
Soleil, afin de d\'eterminer leur multiplicit\'e d'une mani\'ere
homog\'ene. La premi\'ere \'epoque d'observations a fournit 195
candidats companions autour de 117 \'etoiles cibles. La
deuxi\'eme \'epoque d'observations est en cours.
}

\abstract{The multiplicity of early-type stars is still not
well established. The studies of individual star forming
regions suggest a connection with the age and the environment.
To fill in this gap, we started the first detailed
adaptive-optic-assisted imaging survey of 308 BA-type field
stars within 300\,pc from the Sun, to derive their multiplicity
in a homogeneous way.
Our first epoch observations yield 195 companion candidates
around 117 sample targets. The second epoch observations are
underway.
}

\motscles{
\'etoiles: binaires: g\'en\'eral, \'etoiles: binaires:
binaires visuelles, \'etoiles: \'etoiles pr\'ecoces}


\begin{document}

%\maketitle % Exemple de 1ere page sans les valeurs de champs
            % pour afficher les commandes �utiliser

\maketitlepage

\section{Introduction}

The binarity fraction (BF) of early type stars is poorly known
because traditional spectroscopic searches are undermined by
their wide spectral lines. Furthermore, there is evidence that
the BF depends on the stellar density. However, the cluster
multiplicity studies carried out so far can cover only a
limited range of density and age. This prompted us to estimate
the binarity fraction of a representative, volume-limited
sample of early-type field stars.

We designed a survey able to detect at $\sim$10$\sigma$
level an M4-type companion at the mean distance of our sample
($\sim$200\,pc) down to 0.4\,arcsec separation from 100\,Myr
old A-type primary. The companions around B-type stars will be
younger ($\sim$10\,Myr), brighter, and easier to detect. Most
importantly, the physical nature of the candidate companions
is verified by their common proper motion. Our goal is to
compare the properties such as BF and mass ratio of the
multiple stars in the field and in different star-forming
regions. The target accuracy for the BF is 3-5\%.

\section{Sample Selection}

The sample stars were selected according to the following
criteria:

\begin{itemize}
\item spectral types - only BA; most of the stars are B8-A0
because bluer stars are rare and too bright, while redder
stars become too faint to make it into the sample
\item field stars only - the known OB-association members
listed in de Zeeuw et al. (1999) were excluded from the
sample
\item distance $\leq$300\,pc from the Sun as measured by
HIPPARCOS; at the maximum distance the telescope's diffraction
limit of 0.07\,arcsec corresponds to $\sim$21\,A.U. (for
comparison Shatsky \& Tokovinin 2002 probed separations of
45-900\,A.U.)
\item proper motions $>$27\,mas\,yr$^{-1}$ allowing us to
confirm physical companion candidates taking observations at
two epoch separated by 1-2\,yr
\item apparent $V$=5-6\,mag, so the targets are suitable NACO
reference sources even under poor weather, and at the same
time they do not saturate the detector
\item Dec$\leq$+15\,deg, i.e. the targets are visible from the
VLT
\end{itemize}

The final sample consists of 308 stars (136 of them are known
binaries from the Washington catalog), the average distance
is 114\,pc and the median distance is 104\,pc.

\section{Observations}

The observations were carried out with NAOS--CONICA (Nasmyth
Adaptive Optics System -- Near-Infrared Imager and Spectrograph)
at the ESO VLT over the last two years. The pixel scale was
27.03\,mas\,px$^{-1}$, giving 27.7$\times$27.7 arcsec field
of view. Each target was observed at 9 different position on the
detector, collecting total of $\sim$7.5\,min of integration in
$K_S$ or in the intermediate band filter IB\_2.18.

The data reduction includes sky subtraction, flat-fielding,
aligning and combination of the images into a single frame.

\section{Current Status}

As of Mar 2006 we have observed 257 objects from our sample.
We carried out a visual inspection of the combined frames
(with and without PSF subtraction of the target stars), and
{\bf we found 195 companion candidates around 117 sample targets}.
The second epoch observations with $\geq$2\,yr baseline of the
first 16 targets started during ESO Period 77.

\section{Analysis: Modeling the Survey}

To estimate the sensitivity and the completeness of the survey we
have created a Monte-Carlo simulation that takes into account all
available information for the survey stars. The model input
parameters are:

\begin{itemize}
\item the known distances, spectral types and absolute
luminosities for all primaries
\item adopted binarity fraction of 30\%
\item secondary star mass - randomly sampled from the Kroupa
IMF; preferences in the mass ratio of the two components will
be included in the future
\item secondary star's spectral type and absolute magnitude -
calculated from the mass
\item orbital periods - randomly generated from the Duquennoy \&
Mayor (1991) distribution
\item major sxis - calculated from the Kepler's law and the
period
\item random ellipticity and random orbital inclination
\item visibility criterion based on the magnitude difference and
the angular separation between the primary and the companion -
based on the available observations
\end{itemize}

The model predicts: the distributions of periods, angular
separations, magnitude differences and spectral types for the
detected binaries. The simulations indicate that we will detect
$\sim$2/3 of the physical companions.

\acknowledgements
We are grateful to our colleagues from the ESO-Paranal Science
Operations Department who carried out these observation in
Service Mode.

\logbook{09/05/2006}{09/05/2006}{...}

\adressehermes

\end{document}